\documentclass[a4paper, showpacs,12pt]{revtex4}
\usepackage{graphicx}
\usepackage{dcolumn}
\usepackage{bm}
\usepackage{array}
\usepackage{graphicx}
\usepackage{caption2}
\usepackage{amsfonts,amssymb}
\usepackage{amsmath}
\usepackage{indentfirst}

\begin{document}

\title{Controlling transfer of quantum correlations among bi-partitions of a composite quantum system
by combining noisy environments}
\author{Xiu-xing Zhang and Fu-li Li\footnote[2]{Email: flli@mail.xjtu.edu.cn}}
\affiliation{Department of Applied Physics Xi'an Jiaotong
University, Xi'an 710049, China}

\begin{abstract}
The correlation dynamics is investigated for various bi-partitions
of a composite system consisting of two qubits, and two independent
and non-identical noisy environments. The two qubits have no direct
interaction with each other and locally interact with their
environments. Classical and quantum correlations including
entanglement are initially prepared only between the two qubits. We
find that, contrary to the identical noisy environment case, the
entanglement and quantum correlation transfer directions can be
controlled by combining different noisy environments. The amplitude
damping environment determines whether there exists entanglement
transfer among the bi-partitions of a composite system. When one
qubit is coupled to an amplitude damping environment but another one
to a bit-flip one, we find a very interesting result that all the
quantum and classical correlations, and even the entanglement,
originally existing between the qubits, can be completely
transferred without any loss to the qubit coupled to the bit-flip
environment and the amplitude-damping environment. We also notice
that it is possible to distinguish the quantum correlation from the
classical correlation and entanglement by combining different noisy
environments.
\end{abstract}

 \maketitle

\section{Introduction}

Entanglement is a powerful resource for many quantum information
protocols \cite{Nielsen}. It is once considered as a unique
nonclassical correlation existing in a multi-partite quantum state.
However, some investigations have shown that even unentangled states
can also possess nonclassical correlations~\cite{Ollivier,
Henderson, Vedral,ex1}. Computers based on these states can be used
to solve certain tasks more efficiently than their classical
counterparts \cite{Knill}. Besides, distinguishing classical and
quantum correlations in a multi-partite quantum state is vital in
quantum information theory. As a consequence,  a lot of interest
have been devoted to the definition and study of correlations in
quantum systems \cite{Henderson, measure2, measure3, measure4,
measure5}. Among them, quantifying the quantumness of correlations
with quantum discord (QD) \cite{Ollivier}, proposed firstly by
Olliver and Zurek, has received great attention. And till now many
aspects concerning with QD have been discussed, e.g., finding
analytical expression of QD for a certain class of states \cite%
{measure4,analytical,analytical2}, the interplay of QD with quantum
phase transitions \cite {QPT1, QPT2, QPT3, QPT4,QPT5}, studying QD
in continuous variable systems \cite{variable1, variable2}.

As all realistic systems are inevitably coupled to their surrounding
environments which lead to the degradation of their quantum
correlations, recently, the investigation of correlation dynamics
under the influence of various kinds of decoherence scenarios has
become an active subject in both theory and experiment
\cite{Markovian, non-Markovian1, non-Markovian2, sudden tran 1,
sudden tran 2, Quantum dot, Main Ref,ex2,ex3,ex4}. Werlang
\textit{et al.}~\cite{Markovian, non-Markovian1} investigated the QD
dynamics of open systems. It is found that the QD and entanglement
show very different dynamical behaviors and the QD is more robust to
decoherence. Recently, an interesting dynamical behavior of
correlations, named sudden transition, has been observed
\cite{sudden tran 1, sudden tran 2}. It shows that under certain
initial states the correlation undergoes sudden change between a
``classical decoherence" phase and a ``quantum decoherence" phase
\cite {non-Markovian2, sudden tran 1}. This sudden transition
behavior can be explained by a geometrical way and has connection
with the property of environment \cite{non-Markovian2, sudden tran
1}. Moreover, the dynamical behavior of QD is also investigated in
the solid state system \cite{Quantum dot}.

Experimental investigations of nonclassical correlations have also
received much attention \cite{ex1, ex2, ex3, ex4}. Using an optical
architecture, Lanyon \textit{et al.}~\cite{ex1} experimentally
proved that even fully separable states can be used to construct
quantum computer owing to the quantum correlation that contained in
these states. Xu \textit{et al.}~\cite{ex2, ex3} investigated both
the Markovian and non-Markovian dynamics of classical and quantum
correlations and observed the sudden transition behavior of quantum
discord. In a recent paper, Soares-Pinto \textit{et al}.~\cite{ex4}
reported a theoretical and experimental result on the dynamics of
quantum and classical correlations in an NMR quadrupolar system.

We notice that existing investigations on the correlation dynamics
in quantum states consider a total system consisting of two quantum
subsystems and two local environments \cite{Markovian,
non-Markovian1, non-Markovian2, sudden tran 1, sudden tran 2,
Quantum dot, Main Ref,ex2,ex3,ex4}. In general, the quantum systems
are postulated to locally interact with either a common or two
identical environments. However, in practice, two quantum subsystems
may be spatially far away and locally interact with different kinds
of environments. For example, subsystem (A) may be coupled to an
amplitude-damping environment but subsystem (B) may be to a
phase-damping one. In this situation, the question is raised how
correlation dynamics of the subsystems is affected by the different
combination of the environments. In fact, it has been shown that the
correlation dynamics under the action of amplitude-damping reservoir
displays much different behavior in comparison with the action of
phase-damping one \cite{Main Ref}. On the other hand, it may be easy
in experiment to prepare quantum correlations such as entanglement
between certain bi-partitions of a composite system. However,
computation and information processing may require quantum
correlations between other bi-partitions. Therefore, one may need to
transfer quantum correlations initially prepared from one
bi-partition to another. Motivated by these considerations, we here
study the correlation dynamics for various bi-partitions of a
composite system consisting of two qubits $A$ and $B$, and two local
environments $E_{A}$ and $E_{B}$. In contrast to previous
investigations \cite{Markovian, non-Markovian1, non-Markovian2,
sudden tran 1, sudden tran 2, Quantum dot, Main Ref}, we here assume
that $E_{A}$ and $E_{B}$ are non-identical. We find that, contrary
to what is usually stated in the literatures \cite{Main Ref, PRL},
the transfer direction of either the entanglement or other quantum
correlations can be controlled by combining different noisy
environments. We also notice that the amplitude damping process
determines whether there exists entanglement transfer among the
bi-partitions of a composite system. For the amplitude damping
environment plus the bit-flip environment case, we find a very
interesting result that all the quantum and classical correlations,
and even the entanglement, originally existing between the qubits,
can be completely transferred without any loss to the qubit coupled
to the bit-flip environment and the amplitude-damping environment.
We also notice that it is possible to distinguish the quantum
correlation from the classical correlation and entanglement by
combing different environments.

This paper is organized as follows. In Sec. II, several definitions
of quantum and classical correlations are reviewed. In Sec. III, the
model is introduced, and the numerical results are shown and
detailed discussions are given. Finally, a brief summary is given in
Sec. IV.

\section{Classical and Quantum Correlations}

As well known, the total correlation between two random variables
$A$ and$\ B$ can be described by the mutual information
\cite{Nielsen, mutual1, mutual2}. In classical information theory
(CIT), there exist two equivalent expressions for the classical
mutual information \cite{Nielsen, Ollivier, Henderson}
\begin{equation}
I_{C}\left( A\colon B\right) =H(A)+H(B)-H(A,B)  \label{CI1}
\end{equation}%
and%
\begin{equation}
J_{C}\left( A\colon B\right) =H(A)-H(A\Vert B).  \label{CI2}
\end{equation}%
In Eqs. (\ref{CI1}) and (\ref{CI2}) $H(X)=-\sum_{x}p_{x}\log
_{2}p_{x}$ $(X=A$, $B$ and $AB)$ is the Shannon entropy for the
variable $X$ with $p_{x}$ being the probability of $X$ assuming the
value $x$. In Eq. (\ref{CI2}), $H(A\Vert B)=-\sum_{a,b}p_{ab}\log
_{2}p_{a\mid b}=H(A,B)-H(B)$ $(p_{a\mid b}=p_{ab}/p_{b})$ is the
conditional entropy, which represents a weighted average of the
entropies of $A$ given the value of $B$.

In quantum information theory (QIT), the total correlation of a
bipartite system can be expressed in terms of the quantum mutual
information \cite {Nielsen, mutual1, mutual2}
\begin{equation}
I_{q}\left( \rho _{A\colon B}\right) =S(\rho _{A})+S(\rho
_{B})-S(\rho _{AB}),  \label{TC}
\end{equation}%
which is the straightforward extension of (\ref{CI1}). Here, $S(\rho
_{A(B)})=-Tr(\rho _{A(B)}\log _{2}\rho _{A(B)})$ is the von Neumann
entropy of the subsystem $A(B)$, and $S(\rho _{AB})=-Tr(\rho
_{AB}\log _{2}\rho _{AB})$ is the entropy of the composite system
$AB$.

From Eq. (\ref{CI2}), we see that the value of $H(A\Vert B)$ is a
measurement dependence. It is well known that quantum measurement
may fully destroy a quantum state, so the extension of (\ref{CI2})
to quantum realm is no longer straightforward. The counterpart of
(\ref{CI2}) in QIT is defined as
\begin{equation}
J_{q}\left( \rho _{A\colon B}\right) =S(\rho _{A})-S_{\left\{ \Pi
_{j}^{B}\right\} }\left( \rho _{A\mid B}\right) ,  \label{QI2}
\end{equation}%
in which, $\left\{ \Pi _{j}^{B}\right\} $\ describes a set of
projectors performed locally on $B$. $S_{\left\{ \Pi
_{j}^{B}\right\} }\left( \rho _{A\mid B}\right) =\sum_{j}q_{j}S(\rho
_{A}^{j})$ with $\rho _{A}^{j}=\Pi _{j}^{B}\rho _{AB}\Pi
_{j}^{B}/q_{j}$ and the probability $q_{j}=Tr(\Pi _{j}^{B}\rho
_{AB}\Pi _{j}^{B})$. In this article, we choose $\Pi
_{j}^{B}=\left\vert \theta _{j}\right\rangle \left\langle \theta
_{j}\right\vert$, in which $\left\vert \theta _{1}\right\rangle
=\cos \theta \left\vert 0\right\rangle +e^{i\phi }\sin \theta
\left\vert 1\right\rangle$ and $\left\vert \theta _{2}\right\rangle
=-\cos \theta \left\vert1\right\rangle +e^{-i\phi }\sin \theta
\left\vert 0\right\rangle $, with $0\leq \theta \leq \pi $ and
$0\leq \phi \leq 2\pi $. From (\ref{QI2}), we see that different
choices of $\left\{ \Pi _{j}^{B}\right\} $ may lead to different
values of $ J_{q}\left( \rho _{A\colon B}\right) $. The minimum
difference between $ I_{q}\left( \rho _{A\colon B}\right) $ and
$J_{q}\left( \rho _{A\colon B}\right) $, called quantum discord (QD)
\cite{Ollivier}, can be used to describe the quantum correlation of
a bipartite quantum system
\begin{equation}
D\left( \rho _{A\colon B}\right) =\min_{\left\{ \Pi _{j}^{B}\right\}
}\left[ I_{q}\left( \rho _{A\colon B}\right) -J_{q}\left( \rho
_{A\colon B}\right) \right]  \label{QD1}
\end{equation}
or equivalently
\begin{equation}
D\left( \rho _{A\colon B}\right) =I_{q}\left( \rho _{A\colon
B}\right) -\max_{\left\{ \Pi _{j}^{B}\right\} }\left[ J_{q}\left(
\rho _{A\colon B}\right) \right].  \label{QD2}
\end{equation}

Based on (\ref{TC}) and (\ref{QD2}), the classical correlation
contained in a quantum system may be defined as \cite{Henderson}
\begin{equation}
C\left( \rho _{AB}\right) \equiv I_{q}\left( \rho _{A\colon
B}\right) -D\left( \rho _{A\colon B}\right) =\max_{\left\{ \Pi
_{j}^{B}\right\} }\left[ S\left( \rho _{A}\right) -S_{\left\{ \Pi
_{j}^{B}\right\} }\left( \rho _{A\mid B}\right) \right]. \label{CC}
\end{equation}%
Here, we must note that the definitions of (\ref{QD2}) and
(\ref{CC}) valid only when $D\left( \rho _{A\colon B}\right) $ and
$C\left( \rho _{AB}\right) $ are symmetric under the interchange of
$A\leftrightarrow B$. Otherwise, we have to adopt the ``two-side"
measurements for these correlations \cite{two1, two2, two3, two4}.
Under this circumstance, the classical correlation equals to the
maximum classical mutual information that is obtained by local
measurements over both partitions of the system \cite{two1, two2,
two3, two4}
\begin{equation}
K\left( \rho _{AB}\right) \equiv \max_{\left\{ \Pi _{i}^{A}\otimes
\Pi _{j}^{B}\right\} }\left[ I_{C}\left( A\colon B\right) \right] ,
\label{TSC}
\end{equation}%
where $\left\{ \Pi _{i}^{A}\otimes \Pi _{j}^{B}\right\} $ denotes
the set of local measures over subsystems $A$ and $B$, with
$I_{C}\left( A\colon B\right) $ is given by Eq. (\ref{CI1}).

Accordingly, the quantum correlation can be defined as \cite{two1,
two2, two3, two4}%
\begin{equation}
Q\left( \rho _{AB}\right) \equiv I_{q}\left( \rho _{A\colon
B}\right) -K\left( \rho _{AB}\right).  \label{TSQ}
\end{equation}%
Since the two subsystems under consideration are coupled to
non-identical environments and then their states  are asymmetric
about an interchange between the two partitions, we will use Eqs.
(\ref{TSC}) and (\ref{TSQ}) to qualify the classical and quantum
correlations, respectively.

\section{Dynamics of classical and quantum correlations under the action of various combinations of environments}

In this section, we investigate the dynamics of correlations between
various bi-partitions of a system consisting of two qubits and two
independent noisy environments. In contrast to the previously
studied cases where two qubits are coupled to the same kind of noise
environments \cite{Main Ref}, we here consider qubits $A$ and $B$
coupled to independent and non-identical environments. For example,
qubit $A$ is coupled to the amplitude-damping environment but qubit
$B$ to the phase-damping one. We will consider three kinds of
combined noisy environments, i.e., the amplitude-damping plus the
phase-damping environments (APE), the amplitude-damping plus the
bit-flip environments (ABE) and the phase-damping plus the
phase-flip environments (PPE). In the present investigation, it is
always assumed that there is no direct interactions between either
the qubits or the environments.

The initial state of the two qubits $A$ and $B$ is given by
\begin{equation}
\rho _{AB}(0)=\frac{1}{4}\left( \sum_{i=0}^{3}c_{i}\sigma
_{A}^{i}\otimes \sigma _{B}^{i}\right),  \label{initial1}
\end{equation}
where $\sigma _{A,B}^{i}$ is the Pauli matrix $(i=x,y,z)$, $\sigma
_{A,B}^{0}$ is the $2\times 2$ identity matrix, and the coefficients
$c_{i}$ $(c_{0}\equiv 1)$ take real values under the restriction
that $\rho _{AB}$ must be positive and normalized. The states
(\ref{initial1}) represent a general class of quantum states
including both the Werner states $(\left\vert c_{1}\right\vert
=\left\vert c_{2}\right\vert =\left\vert c_{3}\right\vert =a,$
$0<a<1)$ and the Bell states $(\left\vert c_{1}\right\vert
=\left\vert c_{2}\right\vert =\left\vert c_{3}\right\vert =1)$. In
the following discussions, we will set $c_{1}=c_{2}=c_{3}=-a$.

\subsection{Combining Amplitude-Damping and Phase-Damping Environments}

In this subsection, we will investigate the dynamics of correlations
among qubits $A$, $B$, amplitude-damping and phase-damping noisy
environments.

The state map that describes the action of amplitude-damping
environment over qubit $A$ is given by \cite{Main Ref, map1, map2}%
\begin{equation}
\left\vert 0\right\rangle _{A}\left\vert 0\right\rangle
_{E_{A}}\rightarrow \left\vert 0\right\rangle _{A}\left\vert
0\right\rangle _{E_{A}},  \label{Amp1}
\end{equation}
\begin{equation}
\left\vert 1\right\rangle _{A}\left\vert 0\right\rangle
_{E_{A}}\rightarrow
\sqrt{q}\left\vert 1\right\rangle _{A}\left\vert 0\right\rangle _{E_{A}}+\sqrt{p}%
\left\vert 0\right\rangle _{A}\left\vert 1\right\rangle _{E_{A}}.
\label{Amp2}
\end{equation}
where $\left\vert 0\right\rangle _{A}$ and $\left\vert
1\right\rangle _{A}$ are the ground and excited states of qubit $A$,
respectively. $\left\vert 0\right\rangle _{E_{A}}$ and $\left\vert
1\right\rangle _{E_{A}}$ are the states of the environment with no
excitation and one excitation distributed over all its modes,
respectively. Meanwhile, we set $p\in \lbrack 0,1]$ which
corresponds to the reduced parameter of time and $q=1-p$. The
advantage of using $p$ instead of an explicit function of time is
the possibility of describing a wide range of physical systems in
the same model \cite{Nielsen, Main Ref}. The Kraus operators
corresponding to (\ref{Amp1}) and (\ref{Amp2}) are \cite {Nielsen}
\begin{equation}
\Gamma _{1}^{A}=\left(
\begin{array}{cc}
1 & 0 \\
0 & \sqrt{q}%
\end{array}%
\right) ,\Gamma _{2}^{A}=\left(
\begin{array}{cc}
0 & \sqrt{p} \\
0 & 0%
\end{array}%
\right) .  \label{KA1}
\end{equation}

Similarly, the action of phase-damping over qubit $B$ can be
described by the state map \cite{Main Ref, map1, map2}
\begin{equation}
\left\vert 0\right\rangle _{B}\left\vert 0\right\rangle
_{E_{B}}\rightarrow \left\vert 0\right\rangle _{B}\left\vert
0\right\rangle _{E_{B}},  \label{PD1}
\end{equation}%
\begin{equation}
\left\vert 1\right\rangle _{B}\left\vert 0\right\rangle
_{E_{B}}\rightarrow
\sqrt{q}\left\vert 1\right\rangle _{B}\left\vert 0\right\rangle _{E_{B}}+\sqrt{p}%
\left\vert 1\right\rangle _{B}\left\vert 1\right\rangle _{E_{B}}.
\label{PD2}
\end{equation}

The Kraus operators describing the phase-damping are given by \cite{Nielsen}%
\begin{equation}
\Gamma _{1}^{B}=\left(
\begin{array}{cc}
1 & 0 \\
0 & \sqrt{q}%
\end{array}%
\right) ,\Gamma _{2}^{B}=\left(
\begin{array}{cc}
0 & 0 \\
0 & \sqrt{p}%
\end{array}%
\right).  \label{KP1}
\end{equation}

The initial state of the two environments is supposed to be in the
vacuum $\left\vert 00\right\rangle _{E_{A}E_{B}}$. Combined with
(\ref{initial1}), the initial state of the whole system (qubits plus
environments) can be written as
\begin{equation}
\rho _{ABE_{A}E_{B}}(0)=\frac{1}{4}\left( \sum_{i=0}^{3}c_{i}\sigma
_{i}^{A}\otimes \sigma _{i}^{B}\right) \otimes \left\vert
00\right\rangle _{E_{A}E_{B}}\left\langle 00\right\vert .
\label{initial}
\end{equation}

By use of Eqs. (\ref{KA1}), (\ref{KP1}) and (\ref{initial}), the
density matrix of the whole system after the initial moment can be
written as
\begin{equation}
\rho _{ABE_{A}E_{B}}=
\sum_{i,j=1}^{2}\Gamma_i^{(A)}\Gamma_j^{(B)}\rho
_{ABE_{A}E_{B}}(0)\Gamma_i^{(A)\dagger}\Gamma_j^{(B)\dagger}.
\end{equation}
Since we are interested in the correlation dynamics of various
bi-partitions of the whole system, the reduced density matrices are
sufficient for our purpose.

Taking the partial trace of $\rho _{ABE_{A}E_{B}}$ over all
irrelative variables, we can obtain the reduced density matrices for
the various bi-partitions of system. In the representation expanded
by the basis $\left\{ \left\vert 00\right\rangle _{AB},\left\vert
01\right\rangle _{AB},\left\vert 10\right\rangle _{AB},\left\vert
11\right\rangle _{AB}\right\} $, we have the following results. The
reduced density matrix of the qubits A and B is given by
\begin{equation}
\rho _{AB}=\frac{1}{4}\left(
\begin{array}{cccc}
a_{+}+pa_{-} & 0 & 0 & qb_{-} \\
0 & a_{-}+pa_{+} & qb_{+} & 0 \\
0 & qb_{+} & qa_{-} & 0 \\
qb_{-} & 0 & 0 & qa_{+}%
\end{array}%
\right) ,  \label{AB1}
\end{equation}%
where $a_{\pm }=\left( c_{0}\pm c_{3}\right) ,b_{\pm }=\left(
c_{1}\pm c_{2}\right) $.

The reduced density matrices for other bi-partitions
$AE_{A},BE_{B},AE_{B}$ and $BE_{A}$ are as follows
\begin{equation}
\rho _{AE_{A}}=\frac{1}{2}\left(
\begin{array}{cccc}
1 & 0 & 0 & 0 \\
0 & p & \sqrt{pq} & 0 \\
0 & \sqrt{pq} & q & 0 \\
0 & 0 & 0 & 0%
\end{array}%
\right) ,  \label{AEA1}
\end{equation}%
\begin{equation}
\rho _{BE_{B}}=\frac{1}{2}\left(
\begin{array}{cccc}
1 & 0 & 0 & 0 \\
0 & 0 & 0 & 0 \\
0 & 0 & q & \sqrt{pq} \\
0 & 0 & \sqrt{pq} & p%
\end{array}%
\right) ,  \label{BEB1}
\end{equation}%
\begin{equation}
\rho _{AE_{B}}=\frac{1}{4}\left(
\begin{array}{cccc}
1+pqa_{+} & \sqrt{pq}(pa_{+}+a_{-}) & 0 & 0 \\
\sqrt{pq}(pa_{+}+a_{-}) & p(pa_{+}+a_{-}) & 0 & 0 \\
0 & 0 & q(qa_{+}+a_{-}) & q\sqrt{pq}a_{+} \\
0 & 0 & q\sqrt{pq}a_{+} & pqa_{+}%
\end{array}%
\right) ,  \label{AEB1}
\end{equation}%
\begin{equation}
\rho _{BE_{A}}=\frac{1}{4}\left(
\begin{array}{cccc}
a_{+}+qa_{-} & 0 & 0 & \sqrt{pq}b_{-} \\
0 & pa_{-} & \sqrt{pq}b_{+} & 0 \\
0 & \sqrt{pq}b_{+} & a_{-}+qa_{+} & 0 \\
\sqrt{pq}b_{-} & 0 & 0 & pa_{+}%
\end{array}%
\right) .  \label{BEA1}
\end{equation}%
From (\ref{AEA1}) and (\ref{BEA1}), we see that the reduced density
matrices for the bi-partitions $AE_{A}$ and $BE_{B}\ $ are
independent on $c_{i}$, so the correlation dynamical behavior of the
partitions are not affected by the initial states. The reduced
density matrix for partition $E_{A}E_{B}$ is given by
\begin{equation}
\rho _{E_{A}E_{B}}=\frac{1}{4}\left(
\begin{array}{cccc}
(1+q^{2})a_{+}+2qa_{-} & \sqrt{pq}(a_{-}+qa_{+}) & 0 & 0 \\
\sqrt{pq}(a_{-}+qa_{+}) & p(a_{-}+qa_{+}) & 0 & 0 \\
0 & 0 & p(a_{-}+qa_{+}) & p\sqrt{pq}a_{+} \\
0 & 0 & p\sqrt{pq}a_{+} & p^{2}a_{+}%
\end{array}%
\right) .  \label{EAEB1}
\end{equation}

Explicitly knowing the reduced density matrices for all the
bi-partitions, we are in a position to investigate the dynamical
behaviors of various correlations, i.e., total correlation
(\ref{TC}), classical correlation (\ref{TSC}), quantum correlation
(\ref{TSQ}) and entanglement. To qualify entanglement, we use the
Wootters concurrence \cite{concurrence}, which is defined as $C=\max
\left\{ 0,\sqrt{\lambda _{1}}-\sqrt{\lambda _{2}}-\sqrt{\lambda
_{3}}-\sqrt{\lambda _{4}}\right\} $, where ${\lambda _{i}}$ are the
eigenvalues of the matrix $R=\rho _{AB}\left( \sigma _{A}^{y}\otimes
\sigma _{B}^{y}\right) \rho _{AB}^{\ast }\left( \sigma
_{A}^{y}\otimes \sigma _{B}^{y}\right) $ arranged in decreasing
order of magnitude, $\rho _{AB}^{\ast }$ is the complex conjugate of
$\rho _{AB}$ and $\sigma _{A,B}^{y}$ are the Pauli matrices for
subsystems $A$ and $B$. The concurrence attains its maximum value
$1$ for maximally entangled states and $0$ for separable states.

\begin{figure}[htbp]
\centering
\includegraphics[width=12.0cm]{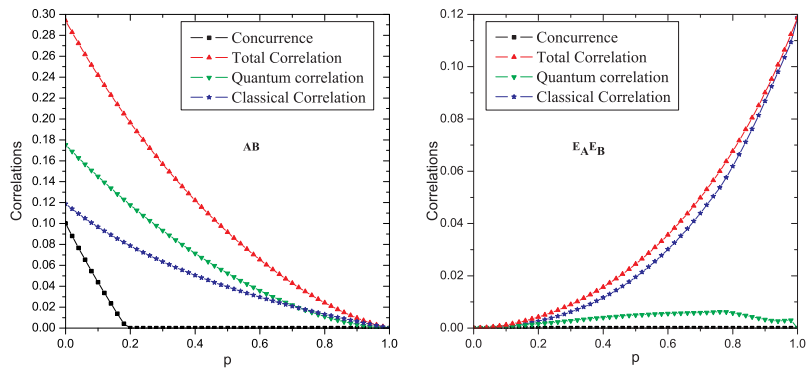}
\caption{Correlation dynamics for the partitions $AB$ and
$E_{A}E_{B}$ under action of the amplitude and phase damping
reservoirs when the initial state is the Werner state ($a=0.4$). The
lines with red triangle, flipped green triangle, blue star and black
square represent total correlation, quantum correlation, classical
correlation and entanglement, respectively. \label{fig:Fig1}}
\end{figure}

In Fig. \ref{fig:Fig1}, various correlations are plotted as a
function of the parameter $p$ for the bi-partitions $AB$ and $
E_{A}E_{B}$ when the initial state is the Werner state with $a=0.4$.
One can see that entanglement appears the sudden death (ESD)
behavior \cite{ESD}. Contrary to what happens to entanglement, the
quantum correlation of $AB$ has no sudden death but monotonously
decreases to zero in the asymptotic limit ($p=1$). It means the
robustness of the quantum correlation to decoherence
\cite{Markovian, non-Markovian1}.

Although the classical correlation between qubits A and B decreases
monotonously as $p$ increases, the classical correlation between
$E_{A}$ and$\ E_{B}$ increases monotonously and arrives at the
maximal value in the asymptotic limit ($p=1$). In this sense, one
may say that the classical correlation lost from the partition $AB$
partially transfers to the partition $E_{A}E_{B}$.

However, much different from the quantum correlation between A and
B, the quantum correlation of the partition $E_{A}E_{B}$ firstly
increases and finally decreases to zero in the asymptotic limit, and
shows the sudden change behavior during the evolution
\cite{analytical}. From Fig. \ref{fig:Fig1}, we also see that there
is no entanglement between $E_{A}$ and$\ E_{B}$ in the whole $p$
range. Therefore, in contrast to the classical correlation, the
entanglement and the quantum correlation for the partition $AB$ are
not transferred to the two environments. This result is much
different from that obtained only with amplitude-damping channels
\cite{Main Ref} and that presented in the reference \cite{PRL}.

\begin{figure}[htbp]
\centering
\includegraphics[width=12.0cm]{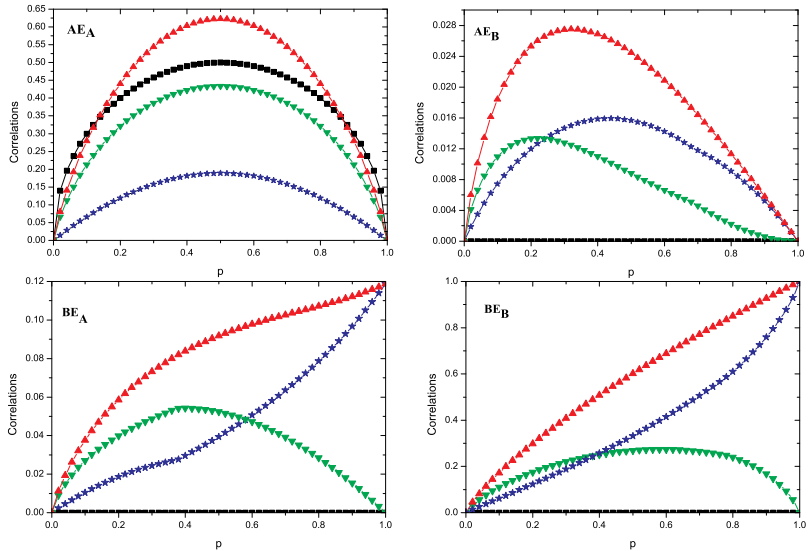}
\caption{Correlation dynamics for the partitions
$AE_{A},AE_{B},BE_{A}$ and $BE_{B}$ under the action of amplitude
and phase damping environments when the initial state is the Werner
state ($a=0.4$). The lines are marked as in Fig. 1.
\label{fig:Fig2}}
\end{figure}

The correlation dynamics for other bi-partitions, i.e.,
$AE_{A},AE_{B},BE_{A}$ and $BE_{B}$, are shown in Fig.
\ref{fig:Fig2}. In these figures we see that the correlation
dynamics between $A$ and the environments differs from that between
$B$ and the environments. All the correlations (classical and
quantum correlations) between $A$ and the environments completely
vanish in the asymptotic limit. However, the classical correlation
can be created between $B$ and $E_{A}$ ($E_{B}$). This result
differs from the situation only with the amplitude-damping
environments, where qubit can only construct nonclassical
correlation with its own environment \cite{Main Ref}. Another
interesting result shown in Fig. \ref{fig:Fig2} is that both the
classical and quantum correlations of $BE_{A}$ exhibit sudden change
behavior \cite{analytical}. Fig. \ref{fig:Fig2} also shows that
during the evolution, the quantum correlation can be constructed in
all the bi-partitions, but entanglement can only be created in
$AE_{A}$.

\begin{figure}[htbp]
\centering
\includegraphics[width=12.0cm]{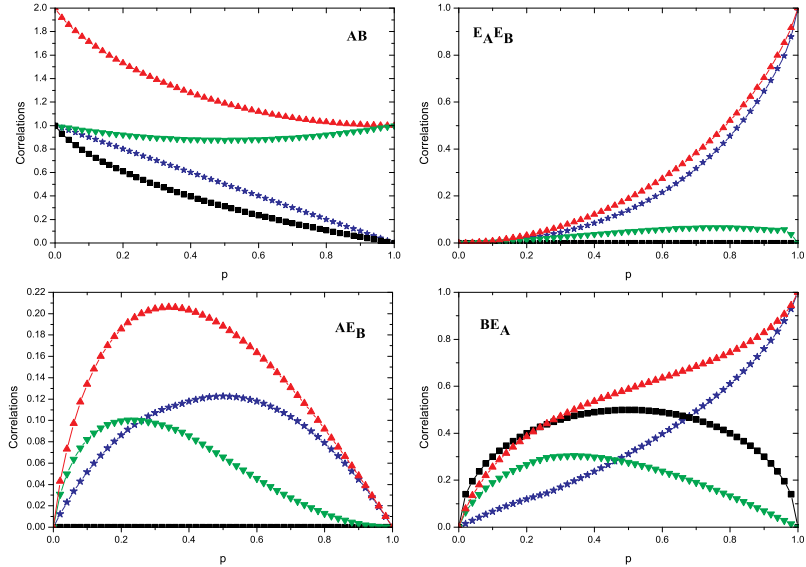}
\caption{Correlation dynamics for the partitions
$AB,E_{A}E_{B},AE_{B}$ and $BE_{A}$ with the amplitude damping and
phase damping environments when the initial state is the Bell state
($a=1$). The lines are marked as in Fig. 1. \label{fig:Fig3}}
\end{figure}

The correlation dynamics for the initial Bell state ($a=1$) is
plotted in Fig. \ref{fig:Fig3}. It is very interesting that in the
asymptotic limit the quantum correlation of $AB$ always remains and
is equal to its initial value $1$ while the entanglement and the
classical correlation decay to zero. This result completely differs
from that of the Werner state as shown in Fig. \ref{fig:Fig2}. It
shows that the nonclassical correlation can exist even in a
separable state \cite{Ollivier}. We also observe that the classical
correlation transfers to the partitions $E_{A}E_{B}$ and $BE_{A}$.

\subsection{Combining Amplitude-Damping and Bit-Flip Environments}

Now let us consider the correlation dynamics for various
bi-partitions of the system under the action of the
amplitude-damping and flip reservoirs. Suppose that the qubits $A$
and $B$ are coupled to the amplitude-damping and bit-flip
environments. The effect of a flip environment over qubit $B$ can be
described by the state map \cite{map1,map2}
\begin{equation}
\left\vert 0\right\rangle _{B}\left\vert 0\right\rangle
_{E_{B}}\rightarrow
\sqrt{p}\left\vert 0\right\rangle _{B}\left\vert 0\right\rangle _{E_{B}}+(i)^{k}%
\sqrt{q}\left\vert 1\right\rangle _{B}\left\vert 1\right\rangle
_{E_{B}}, \label{BF1}
\end{equation}%
\begin{equation}
\left\vert 1\right\rangle _{B}\left\vert 0\right\rangle
_{E_{B}}\rightarrow
\sqrt{p}\left\vert 1\right\rangle _{B}\left\vert 0\right\rangle _{E_{B}}+(-i)^{k}%
\sqrt{q}\left\vert 0\right\rangle _{B}\left\vert 1\right\rangle
_{E_{B}}, \label{BF2}
\end{equation}%
where $k=0$ is for the bit-flip environment, $k=1$ for the
bit-phase-flip environment. Since the correlation dynamics does not
depend on $k$, in what follows, we will consider only the bit-flip
environment ($k=0$). The Kraus operators of the bit-flip are as follows \cite{Nielsen}%
\begin{equation}
\Gamma _{1}^{B}=\left(
\begin{array}{cc}
\sqrt{p} & 0 \\
0 & \sqrt{p}%
\end{array}%
\right) ,\Gamma _{2}^{B}=\left(
\begin{array}{cc}
0 & \sqrt{q} \\
\sqrt{q} & 0%
\end{array}%
\right) .  \label{KB1}
\end{equation}

The initial state of the whole system is still supposed to be
(\ref{initial}). From Eqs. (\ref{Amp1}), (\ref{Amp2}), (\ref{BF1})
and (\ref{BF2}), we can obtain the density matrix of the whole
system. The reduced density matrices for various bi-partitions can
then be analytical obtained, which are given in the appendix A.

\begin{figure}[htbp]
\centering
\includegraphics[width=12.0cm]{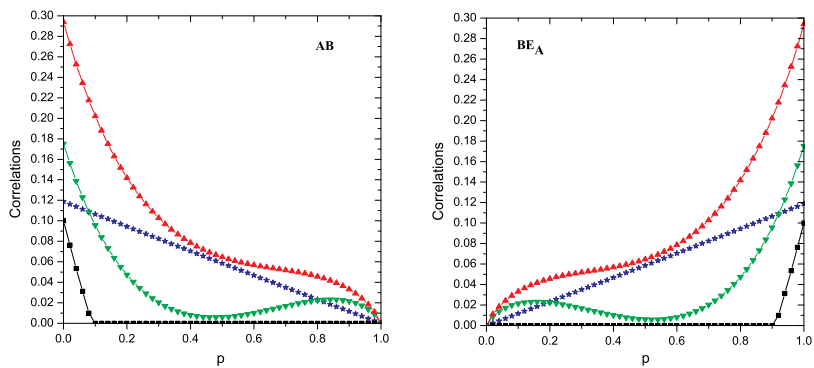}
\caption{Correlation dynamics for the partitions $AB$ and $BE_{A}$
with the combined amplitude damping and bit flip environments when
the initial state is the Werner state ($a=0.4$). The lines are
marked as in Fig. 1. \label{fig:Fig4}}
\end{figure}

\begin{figure}[htbp]
\centering
\includegraphics[width=12.0cm]{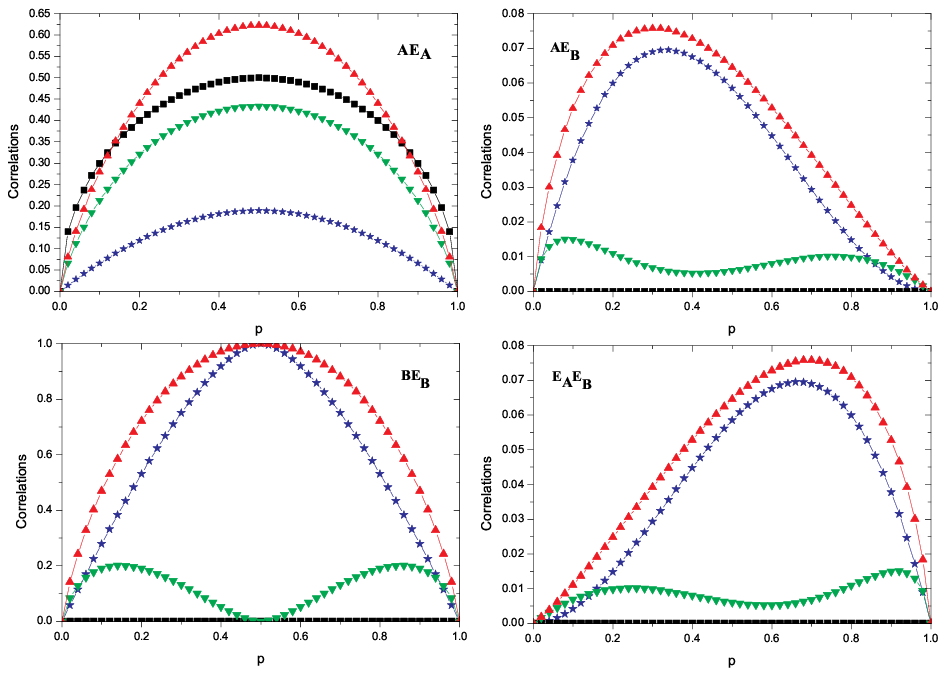}
\caption{Correlation dynamics for the partitions
$AE_{A},AE_{B},BE_{B}$ and $E_{A}E_{B}$ with the combined amplitude
damping and bit fip environments when the initial state is the
Werner state ($a=0.4$). The lines are marked as in Fig. 1.
\label{fig:Fig5}}
\end{figure}

In Fig. \ref{fig:Fig4}, the correlation dynamics for the bipartitions $AB$ and $%
BE_{A}$ are shown. We observe that the total, classical and quantum
correlations vanish at the asymptotic limit $p=1$ but the
entanglement suddenly and completely  disappears at a finite point.
It is also noticed that in the asymptotic limit all the correlations
including entanglement are fully transferred to the partition
$BE_{A}$ rather than that of the two environments $E_{A}$ and
$E_{B}$ \cite {PRL, Main Ref}. In Fig. \ref{fig:Fig5}, the
correlation dynamics for the other four bi-partitions are shown.
From these figures, one may see that although there exist bipartite
correlations during the evolution process, they all vanish in the
asymptotic limit. The four bi-partitions $AE_{A},AE_{B},BE_{B}$ and
$E_{B}E_{A}$ can then be treated as transfer bridges for
correlations.

\begin{figure}[htbp]
\centering
\includegraphics[width=12.0cm]{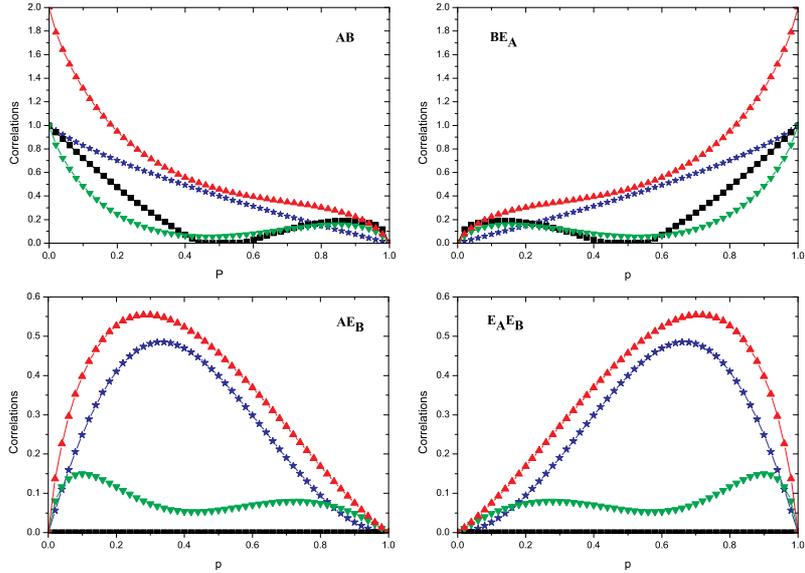}
\caption{Correlation dynamics for the partitions
$AB,E_{A}E_{B},AE_{B}$ and $BE_{A}$ with the combined amplitude
damping and bit flip environments when the initial state is the Bell
state ($a=1$). The lines are marked as in Fig. 1. \label{fig:Fig6}}
\end{figure}

In Fig. \ref{fig:Fig6}, the correlation dynamics for various
partitions of the system with the initial Bell state is shown. It is
noticed that all the correlations and entanglement of $AB$ can be
fully transferred to the partition $BE_{A}$. One may also find that
there is the death and revival of entanglement under the action of
the combined amplitude-damping and bit-flip environments for the
Bell state with $a=1$.

From the above discussions, we obtain a very interesting result that
all the quantum and classical correlations, and even the
entanglement, originally prepared between the qubits $A$ and $B$,
can be completely transferred to the partition $BE_{A}$ under the
action of the amplitude damping and bit-flip environments.

\subsection{Combining Phase-Damping and Phase-Flip Environments}

In this subsection, we will consider the situation where two quits
$A$ and $B$ interact independently and locally with the
phase-damping and phase-flip environments, respectively. The
corresponding state map induced by a phase-flip environment over
qubit $B$ is given by \cite{map1, map2}
\begin{equation}
\left\vert 0\right\rangle _{B}\left\vert 0\right\rangle
_{E_{B}}\rightarrow
\sqrt{p}\left\vert 0\right\rangle _{B}\left\vert 0\right\rangle _{E_{B}}+\sqrt{q}%
\left\vert 0\right\rangle _{B}\left\vert 1\right\rangle _{E_{B}},
\label{PF1}
\end{equation}%
\begin{equation}
\left\vert 1\right\rangle _{B}\left\vert 0\right\rangle
_{E_{B}}\rightarrow
\sqrt{p}\left\vert 1\right\rangle _{B}\left\vert 0\right\rangle _{E_{B}}-\sqrt{q}%
\left\vert 1\right\rangle _{B}\left\vert 1\right\rangle _{E_{B}}.
\label{PF2}
\end{equation}
The Kraus operators of the phase-flip environment
are as follows \cite{Nielsen}%
\begin{equation}
\Gamma _{1}^{B}=\left(
\begin{array}{cc}
\sqrt{p_{1}} & 0 \\
0 & \sqrt{p_{1}}%
\end{array}%
\right) ,\Gamma _{2}^{B}=\left(
\begin{array}{cc}
\sqrt{q_{1}} & 0 \\
0 & -\sqrt{q_{1}}%
\end{array}%
\right) .  \label{KPF1}
\end{equation}

It is known that a unitary recombination of the elements in
(\ref{KP1}) can give the expression of (\ref{KPF1}) \cite{Nielsen}.
In particular, the phase-damping process is exactly the same as the
phase-flip one when $p_{1}=(1+\sqrt{1-q})/2$. In the present
investigation, we set $p_{1}=p$ and $q_{1}=1-p_{1}$ which guarantees
that $A$ and $B$ undergoing non-identical environments.

Starting from the initial state defined by Eq. (\ref{initial}) and
using the same method mentioned in the preceding subsections, we can
obtain the reduced density matrices for various bi-partitions of the
system, which are shown in the appendix B.

\begin{figure}[htbp]
\centering
\includegraphics[width=12.0cm]{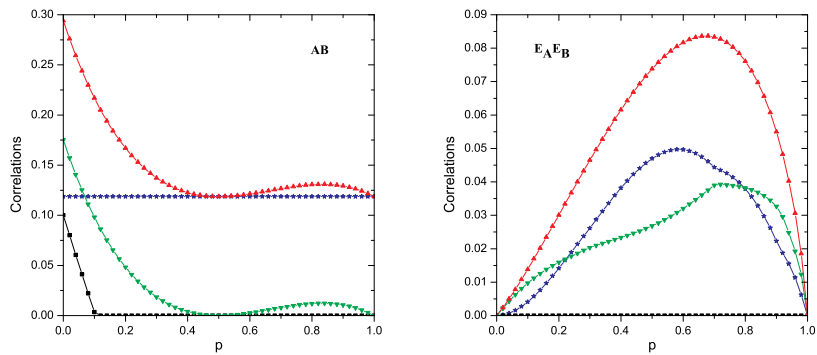}
\caption{Correlation dynamics for the partitions $AB$ and
$E_{A}E_{B}$ with the combined phase damping and phase flip
environments when the initial state is the Werner state ($a=0.4$).
The lines are marked as in Fig. 1. \label{fig:Fig7}}
\end{figure}

\begin{figure}[htbp]
\centering
\includegraphics[width=12.0cm]{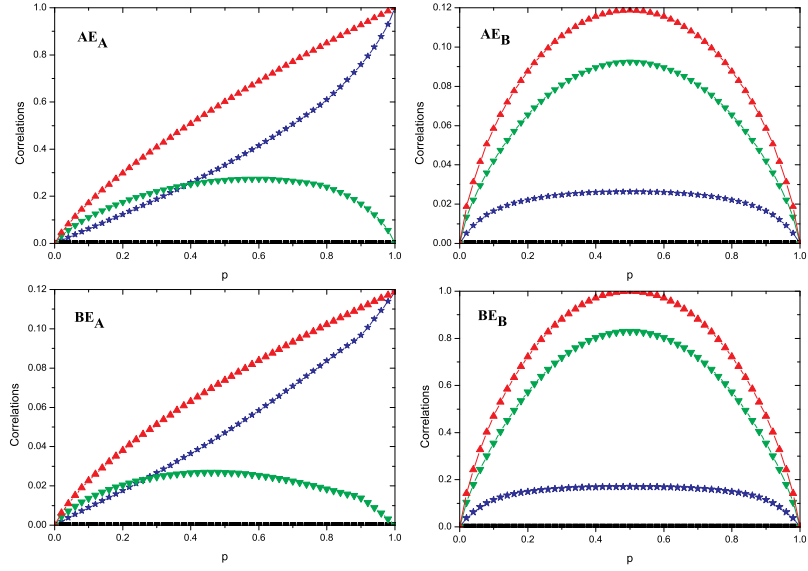}
\caption{Correlation dynamics for the partitions
$AE_{A},AE_{B},BE_{A}$ and $BE_{B}$ with the combined phase damping
and phase flip environments for the Werner state ($a=0.4$). The
lines are marked as in Fig. 1. \label{fig:Fig8}}
\end{figure}

\begin{figure}[htbp]
\centering
\includegraphics[width=12.0cm]{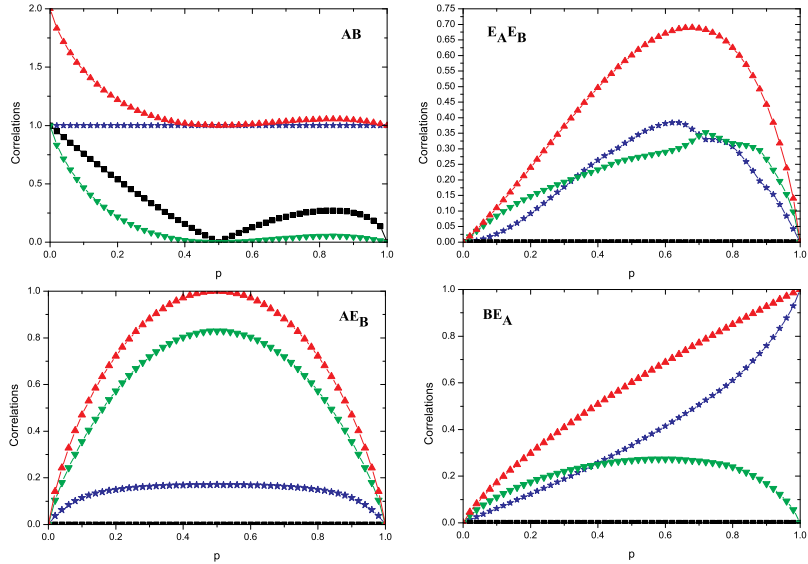}
\caption{Correlation dynamics for the partitions
$AB,E_{A}E_{B},AE_{B}$ and $BE_{A}$ with the combined phase damping
and phase flip environments for the Bell state ($a=1$). The lines
are marked as in Fig. 1. \label{fig:Fig9}}
\end{figure}

In Fig. \ref{fig:Fig7}, the correlation dynamics for the partitions
$AB$ and $E_{A}E_{B}$ is shown. It can be seen that the total and
quantum correlations display the decay and  revival behavior while
the entanglement exhibits the ESD. This result differs from that of
the pure phase-damping environment where the quantum and total
correlations monotonously decrease with $p$ increasing \cite{Main
Ref}. In particular, the classical correlation of $AB$ remains a
constant in the whole range. This indicates that the classical
correlation is immune to the decoherence. This result may lead to an
operational way of measuring both the classical and quantum
correlation in a composite system \cite{analytical}. All the
correlations of $E_{A}E_{B}$ firstly increase and reach to the
maximal values with $p$ increasing, and then decrease to zero in the
asymptotic limit ($p=1$). This phenomenon differs also from that of
the pure phase-damping environment where classical and quantum
correlations between the environments reach to a same finite value
at $p=1$ \cite{Main Ref}.

Fig. \ref{fig:Fig8} shows the correlation evolution for other
bi-partitions of the system. It is observed that only the classical
correlations can be created among the bi-partitions $AE_{A}$ and
$BE_{A}$ in the asymptotic limit ($p=1$). The quantum correlation
can exist among the various bi-partitions only during the evolution.
However, there exist no entanglement between any partitions during
the whole evolution and the entanglement exiting in the initial
state is fully evaporated, which is very similar to that of the pure
phase-damping noisy environment \cite{Main Ref}.

The correlation dynamics is shown in Fig. \ref{fig:Fig9} when the
two qubits are initially in the Bell state . In these figures, it is
seen that the entanglement displays the death and revival phenomenon
and the classical correlation keeps constant during the evolution.
The quantum correlation and entanglement existing in the initial
state are fully evaporated in the asymptotic limit ($p=1$). As the
same as shown in Fig. \ref{fig:Fig8}, only the classical correlation
between the qubit $A(B)$ and the phase-damping environment can be
created.

\section{Summary}

We study the quantum and classical correlation (including
entanglement) dynamics of two qubits $A$ and $B$ which are
independently and locally coupled to two non-identical environments
$E_{A}$ and $E_{B}$, respectively. Three different combinations of
noisy environments are considered, i.e., the amplitude plus phase
damping environments (APE), the amplitude-damping plus the bit-flip
environments (ABE) and the phase-damping plus the phase-flip
environments (PPE). We find that, contrary to the single-type noisy
environment case, the entanglement and quantum correlation transfer
direction can be controlled by combining the different noisy
environments. This may provides us a way of manipulating the
correlation dynamics of a composite system. The amplitude-damping
environment plays a very important role in the correlation dynamics.
It determines whether there exists a correlation transfer among the
bi-partitions of a composite system. If there is no amplitude
damping, e.g. in the PPE, the initial entanglement is completely
evaporated. In the ABE case we show that all the quantum and
classical correlations, and even entanglement, which exist in the
initial state of the two qubits, can be completely transferred to
the bi-partition $BE_{A}$ without any loss. When the initial state
is the Bell one, in the APE case the quantum correlation between the
two qubits can remain the initial value in the asymptotic limit
while the classical correlation and entanglement decay to zero, but
in the PPE case the classical correlation keeps constant while the
quantum correlation and entanglement decay to zero. This result
shows that it is possible to distinguish the quantum correlation
form the classical correlation and entanglement by combining
different kinds of environments. In the combined environments ABE
and PPE, even for the Bell state, the entanglement can display the
sudden death (ESD) and revival phenomenon. We also notice that the
quantum correlation can appear in arbitrary bi-partitions and have
no sudden death behavior during the evolution, which means the
robustness of the quantum correlation to decoherence.

\begin{acknowledgments}
This work is supported by the National Basic Research Program of
China (973 Program) No. 2010CB923102, and the National Nature
Science Foundation of China (Grand No. 60778021).
\end{acknowledgments}

\subsection{Appendix A: Reduced Density Matrices for Various Bi-partitions under the
Action of Amplitude-Damping and Bit-Flip Environments}

The reduced density matrix for the partition $AB$, obtained by
taking the partial trace of $\rho _{ABE_{A}E_{B}}$ over the degrees
of freedom of the
reservoir is given by%
\begin{equation}
\rho _{AB}=\frac{1}{4}\left(
\begin{array}{cccc}
2p^{2}+qa_{-} & 0 & 0 & \sqrt{q}(pb_{-}+qb_{+}) \\
0 & 2p^{2}+qa_{+} & \sqrt{q}(pb_{+}+qb_{-}) & 0 \\
0 & \sqrt{q}(pb_{+}+qb_{-}) & q(pa_{-}+qa_{+}) & 0 \\
\sqrt{q}(pb_{-}+qb_{+}) & 0 & 0 & q(qa_{-}+pa_{+})%
\end{array}%
\right) .  \label{AB2}
\end{equation}%
For the partitions $AE_{A},BE_{B},AE_{B}$ and $BE_{A}$, the
reduced-density
operators are as follows%
\begin{equation}
\rho _{AE_{A}}=\frac{1}{2}\left(
\begin{array}{cccc}
1 & 0 & 0 & 0 \\
0 & p & \sqrt{pq} & 0 \\
0 & \sqrt{pq} & q & 0 \\
0 & 0 & 0 & 0%
\end{array}%
\right) ,  \label{AEA2}
\end{equation}%
\begin{equation}
\rho _{BE_{B}}=\frac{1}{2}\left(
\begin{array}{cccc}
p & 0 & 0 & \sqrt{pq} \\
0 & q & \sqrt{pq} & 0 \\
0 & \sqrt{pq} & p & 0 \\
\sqrt{pq} & 0 & 0 & q%
\end{array}%
\right) ,  \label{BEB2}
\end{equation}%
\begin{equation}
\rho _{AE_{B}}=\frac{1}{2}\left(
\begin{array}{cccc}
(p+p^{2}) & 0 & 0 & q\sqrt{p}c_{1} \\
0 & (1+p)q & q\sqrt{p}c_{1} & 0 \\
0 & q\sqrt{p}c_{1} & pq & 0 \\
q\sqrt{p}c_{1} & 0 & 0 & q^{2}%
\end{array}%
\right) ,  \label{AEB2}
\end{equation}%
\begin{equation}
\rho _{BE_{A}}=\frac{1}{4}\left(
\begin{array}{cccc}
(1+p)D & 0 & 0 & \sqrt{p}B \\
0 & pC & \sqrt{p}A & 0 \\
0 & \sqrt{p}A & (1+p)C & 0 \\
\sqrt{p}B & 0 & 0 & pD%
\end{array}%
\right) ,  \label{BEA2}
\end{equation}%
where $A=(qb_{-}+pb_{+}),B=(pb_{-}+qb_{+}),C=(pa_{-}+qa_{+})$ and $%
D=(pa_{+}+qa_{-})$.

After tracing out the degrees of freedom of the two qubits, we
obtain the reduced density matrix for the two environments $E_{A}$
and $E_{B}$
\begin{equation}
\rho _{E_{A}E_{B}}=\frac{1}{2}\left(
\begin{array}{cccc}
p(1+q) & 0 & 0 & p\sqrt{q}c_{1} \\
0 & q(2-p) & p\sqrt{q}c_{1} & 0 \\
0 & p\sqrt{q}c_{1} & p^{2} & 0 \\
p\sqrt{q}c_{1} & 0 & 0 & pq%
\end{array}%
\right).  \label{EAEB2}
\end{equation}

\subsection{Appendix B: Density Matrices for Various Bi-partitions under the
Action of Phase-Damping and Phase-Flip Environments}

The reduced density matrix for $AB$ under the influence of the mixed
phase damping and phase flip environments is%
\begin{equation}
\rho _{AB}=\frac{1}{4}\left(
\begin{array}{cccc}
a_{+} & 0 & 0 & \sqrt{q}(p-q)b_{-} \\
0 & a_{-} & \sqrt{q}(p-q)b_{+} & 0 \\
0 & \sqrt{q}(p-q)b_{+} & a_{-} & 0 \\
\sqrt{q}(p-q)b_{-} & 0 & 0 & a_{+}%
\end{array}%
\right) .  \label{AB3}
\end{equation}%
For the partitions $AE_{A},BE_{B},AE_{B}$ and $BE_{A}$, the
reduced-density
operators are analytically given by%
\begin{equation}
\rho _{AE_{A}}=\frac{1}{2}\left(
\begin{array}{cccc}
1 & 0 & 0 & 0 \\
0 & 0 & 0 & 0 \\
0 & 0 & q & \sqrt{pq} \\
0 & 0 & \sqrt{pq} & p%
\end{array}%
\right) ,  \label{AEA3}
\end{equation}%
\begin{equation}
\rho _{BE_{B}}=\frac{1}{2}\left(
\begin{array}{cccc}
p & \sqrt{pq} & 0 & 0 \\
\sqrt{pq} & q & 0 & 0 \\
0 & 0 & p & -\sqrt{pq} \\
0 & 0 & -\sqrt{pq} & q%
\end{array}%
\right) ,  \label{BEB3}
\end{equation}%
\begin{equation}
\rho _{AE_{B}}=\frac{1}{2}\left(
\begin{array}{cccc}
p & \sqrt{pq}c_{3} & 0 & 0 \\
\sqrt{pq}c_{3} & q & 0 & 0 \\
0 & 0 & p & -\sqrt{pq}c_{3} \\
0 & 0 & -\sqrt{pq}c_{3} & q%
\end{array}%
\right),  \label{AEB3}
\end{equation}%
\begin{equation}
\rho _{BE_{A}}=\frac{1}{4}\left(
\begin{array}{cccc}
a_{+}+qa_{-} & 0 & \sqrt{pq}a_{-} & 0 \\
0 & a_{-}+qa_{+} & 0 & \sqrt{pq}a_{+} \\
\sqrt{pq}a_{-} & 0 & pa_{-} & 0 \\
0 & \sqrt{pq}a_{+} & 0 & pa_{+}%
\end{array}%
\right) .  \label{BEA3}
\end{equation}%
Similarly, the reduced density matrix for partition $E_{A}E_{B}$ is%
\begin{equation}
\rho _{E_{A}E_{B}}=\frac{1}{2}\left(
\begin{array}{cccc}
p(1+q) & -p\sqrt{pq}c_{3} & p\sqrt{pq} & -pqc_{3} \\
-p\sqrt{pq}c_{3} & q(1+q) & -pqc_{3} & q\sqrt{pq} \\
p\sqrt{pq} & -pqc_{3} & p^{2} & -p\sqrt{pq}c_{3} \\
-pqc_{3} & q\sqrt{pq} & -p\sqrt{pq}c_{3} & pq%
\end{array}%
\right).  \label{EAEB3}
\end{equation}

\bigskip

\end{document}